\documentclass[aps,pre,floatfix]{revtex4}
\usepackage{epsf}
\usepackage{subfigure}
\usepackage{amsmath}
\usepackage{amssymb}
\usepackage{graphicx}
\usepackage{epstopdf}
\begin{document}

\newcommand{\be}{\begin{equation}}
\newcommand{\ee}{\end{equation}}
\newcommand{\bea}{\begin{eqnarray}}
\newcommand{\eea}{\end{eqnarray}}
\newcommand{\mean}[1]{\left \langle #1 \right \rangle}

\title{\bf Dynamical Aspects of Information in Copolymerization Processes}

\author{Pierre Gaspard}
\affiliation{Center for Nonlinear Phenomena and Complex Systems,\\
Universit\'e Libre de Bruxelles, Campus Plaine, Code Postal 231,
B-1050 Brussels, Belgium}

\begin{abstract}
Natural supports of information are given by random copolymers such as DNA or RNA where information is coded in the sequence of  covalent bonds. At the molecular scale, the stochastic growth of a single copolymer with or without a template proceeds by successive random attachments or detachments of monomers continuously supplied by the surrounding solution. The thermodynamics of copolymerization shows that fundamental links already exist between information and thermodynamics at the molecular scale, which opens new perspectives to understand the dynamical aspects of information in biology.  

\vskip 0.3 cm

\end{abstract}

\noindent {Contribution to the Proceedings of ECCS'12, Brussels, September 3-7, 2012}

\vskip 0.5 cm

\maketitle

\section{Introduction}

Under nonequilibrium conditions, the emergence of dynamical order is already in action at the molecular scale during copolymerization processes.  Copolymers are special because they constitute the smallest physico-chemical supports of information.  Little is known about the thermodynamics and kinetics of information processing in copolymerizations although such reactions play an essential role in many complex systems, e.g. in biology.  In this context, recent advances have been performed which shed a new light on the nonequilibrium constraints required to generate information-rich copolymers \cite{AG08,J08,AG09,G12}.

Natural supports of information are given by random copolymers where information is coded in the sequence of covalent bonds, as already suggested by Schr\"odinger with his concept of aperiodic crystal \cite{S44}.    Random copolymers exist in chemical and biological systems.  Examples are styrene-butadiene rubber, proteins, RNA, and DNA, this latter playing the role of information support in biology.

At the molecular scale, the stochastic growth of a single copolymer proceeds by successive random attachments or detachments of monomers $\{m\}$ continuously supplied by the surrounding solution:
\begin{equation}
m_1m_2\cdots m_{l-1} \ + \ m_{l} \ \rightleftharpoons
  \  m_1m_2\cdots m_{l-1}m_{l}
\end{equation}
The solution is supposed to be sufficiently large to play the role of a reservoir where the concentrations of monomers are kept constant.  In this regard, the stochastic growth of a single copolymer is modeled by a Markovian process with transition rates depending on the fixed concentrations of monomers in the surrounding solution \cite{AG08,J08,AG09,G12}.

According to local detailed balancing, the rates of forward and reversed transitions have ratios that are determined by the free energies of the copolymers $m_1m_2\cdots m_{l}$ in physical equilibrium with the surrounding solution.
Thermodynamic quantities can thus be defined and their time evolution studied during the copolymerization process.  
In this way, fundamental relationships can be established between the thermodynamics of copolymerization processes and the information content encoded in growing copolymers \cite{AG08,J08,AG09,G12}.  The purpose of this communication is to present the latest results obtained in this framework.

\section{Results}

As shown in Ref.~\cite{AG08} for copolymerization with or without a template, the thermodynamic entropy production is related not only to the average value of the free energy per monomer in the grown sequence, but also to the Shannon disorder of the sequence itself.  This result is at the origin of dissipation-error tradeoff during copolymer growth \cite{B79}.

Two growth regimes are identified: 

(1) A regime close to the thermodynamic equilibrium where the copolymer can grow in an adverse free-energy landscape by the entropic effect of its Shannon disorder.  In this regime, the disorder of the grown sequence dominates the process even in the presence of a template, in which case the copying process generates a lot of errors.

(2) A regime farther away from equilibrium where the growth proceeds because the free energy of monomer attachment is favorable.  In this regime, the error rate drops to low values.

In Refs.~\cite{AG08,AG09}, these regimes were studied as a function of the free energy driving force. In Ref.~\cite{G12}, results have been reported on a model of free copolymerization where the attachment and detachment rates are controlled by the concentrations of monomers in the surrounding solution.  In the present communication, this study of the dependence on monomeric concentrations is extended to a model of copolymerization with a template. The template as well as the growing copy are composed of two monomers $m=1$ and $m=2$.  The pairs $1$-$1$ and $2$-$2$ are favored between the template $\alpha$ and the copy $\omega$.  The pairs $1$-$2$ and $2$-$1$ are considered as errors during information transmission from the template to the copy.  The kinetic mechanism of elongation is the following:
\begin{equation}
 \begin{array}{l}
 \alpha: \\
\omega:
 \end{array}
\qquad
\begin{array}{l}
n_1\ n_2\; \cdots \, n_{l-1}\, n_{l} \ n_{l+1}\ \cdots\\
m_1m_2\cdots m_{l-1}
\end{array}
 \quad
 \begin{array}{l}
  \\
 + \ m_{l}
 \end{array}
  \quad \rightleftharpoons
  \quad
  \begin{array}{l}
n_1\ n_2\; \cdots \, n_{l-1}\, n_{l} \ n_{l+1}\ \cdots\\
m_1m_2\cdots m_{l-1} m_{l}
\end{array}
\end{equation}
The attachment rates are given by $w_{+m\vert n} = k_{+ m \vert n} [m]$ and the detachment rates by $w_{-m\vert n} = k_{-m \vert n}$.  The attachement rates are proportional to the monomeric concentrations $[m]$, while the detachment rates are not since detachments do not need the presence of monomers in the surrounding solution.  The rates are supposed to be independent of the end $m_{l-1}$ of the copy $\omega$, which is a simplifying assumption.  The rate of formation of correct pairs is defined as $k_{\rm correct}\equiv k_{+1\vert 1}=k_{+2\vert 2}$ and the error rate as $k_{\rm error}\equiv k_{+1\vert 2}=k_{+2\vert 1}$.  All the detachment rates are assumed to take the same value: $k_{\rm off}\equiv k_{-1\vert 1}=k_{-1\vert 2}=k_{-2\vert 1}=k_{-2\vert 2}$.

\begin{figure}
\centering
\includegraphics[height=8cm]{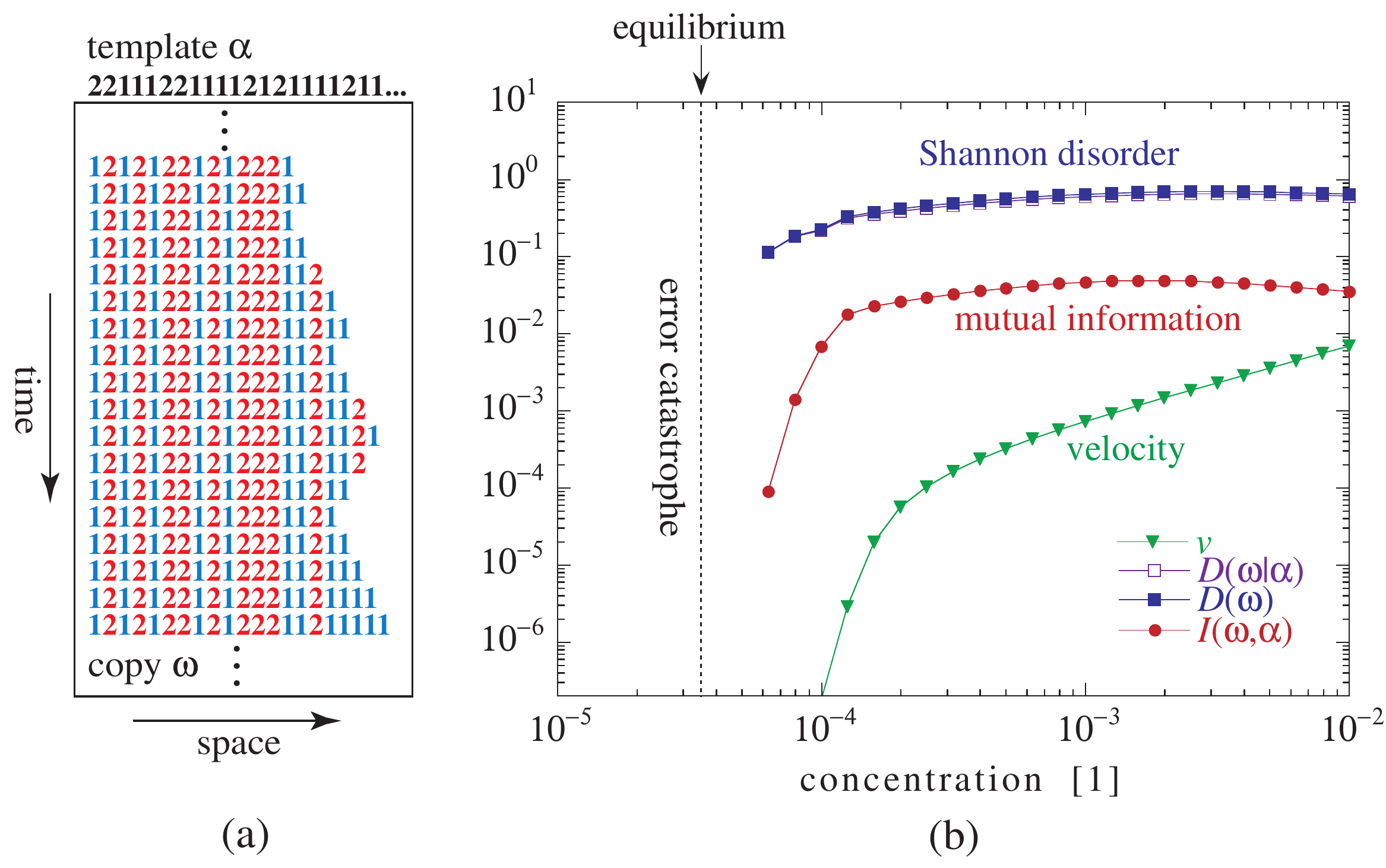}
\caption{Stochastic growth of the copolymer $\omega$ on the template $\alpha$, as simulated by Gillespie's algorithm with the parameter values  $k_{\rm correct}=1$, $k_{\rm error}=0.5$, $k_{\rm off}=10^{-3}$, and $[2]=1.3\times 10^{-3}$. The template is generated by a Bernoulli process of probabilities~$(\frac{1}{2},\frac{1}{2})$.  (a)~Space-time plot at the concentration $[1]=2\times 10^{-3}$; (b)~The mean growth velocity $v$, the Shannon disorder $D(\omega)$ of the copy $\omega$, the Shannon disorder $D(\omega\vert\alpha)$ of the copy $\omega$ conditioned to the template sequence $\alpha$, and the mutual information $I(\omega,\alpha)$ between the copy $\omega$ and the template $\alpha$ versus the concentration $[1]$ of the monomers of species~$1$.}
\label{figure}
\end{figure}

The stochastic process is simulated with Gillespie's algorithm.  Figure \ref{figure}(a) illustrates the fluctuating growth of a copy in space and time.  In this example, the error rate is larger because of the smallness of the ratio $k_{\rm correct}/k_{\rm error}=2$.  This ratio is determined by the strength of the pairing bonds.  Figure~\ref{figure}(b) depicts the mean growth velocity as well as the quantities characterizing the information content of the copy $\omega$ compared to the template $\alpha$ versus the concentration of monomers~$1$ in the surrounding solution.  The mean growth velocity vanishes at equilibrium, which exists at the concentration $[1]_{\rm eq}\simeq 0.3\times 10^{-4}$ if $[2]=1.3\times 10^{-3}$.  Both the Shannon disorder $D(\omega)$ of the copy and the conditional disorder $D(\omega\vert\alpha)$ of the copy with respect to the template are larger than the mutual information $I(\omega,\alpha)$ between the copy and the template.  In any case, the three quantities are related to each other by the well-known formula
\begin{equation}
I(\omega,\alpha) = D(\omega) - D(\omega\vert\alpha)
\end{equation}
from information theory \cite{AG08,AG09}.  The mutual information characterizes the fidelity of information transmission between the template and the copy.  Figure~\ref{figure}(b) shows that this fidelity decreases close to equilibrium.  The reason is that the out-of-equilibrium directionality is lost close to equilibrium where fluctuations go either forward or backward because the principle of detailed balancing prevails at equilibrium.  Consequently, there is a multiplication of errors close to equilibrium.  This error catastrophe is avoided by maintaining the system far enough from equilibrium.

\section{Conclusions and perspectives}

The results show that fidelity in copying a copolymer requires the supply of enough free energy from the attachment of monomers.  In this respect, the nonequilibrium driving should exceed a critical value in order to transmit information in copolymerization processes with a template, such as DNA replication~\cite{AG08}.   The statement by Manfred Eigen that ``information cannot originate in a system that is at equilibrium"\cite{E92} is rigorously proved in the present framework.  The thermodynamics of copolymerization thus shows that fundamental links already exist  at the molecular scale between information and thermodynamics~\cite{AG08,J08,AG09,G12}.

The transition between the two growth regimes could be experimentally investigated in chemical or biological copolymerizations.  In polymer science, methods have not yet been much developed to perform the synthesis and sequencing of copolymers for the information they may support.  However, such methods are already well developed for DNA and under development for single-molecule DNA or RNA sequencing \cite{GB06,Eid09,UEAKFTP10}.  These methods could be used to test experimentally the predictions of copolymerization thermodynamics by varying NTP and pyrophosphate concentrations to approach the regime near equilibrium where the mutation rate increases.  

These considerations open new perspectives to understand the dynamical aspects of information in biology.  During copolymerization processes with a template (as it is the case for replication, transcription or translation in biological systems), information is transmitted although errors may occur due to molecular fluctuations, which are sources of mutations.  The two main features of biological systems -- namely, metabolism and self-reproduction -- turn out to be related in a fundamental way since information processing is constrained by energy dissipation during copolymerizations.  Moreover, the error threshold for the emergence of quasi-species in the hypercycle theory by Eigen and Schuster \cite{ES77-78} could be induced at the molecular scale by the transition towards high fidelity replication beyond the transition between the two growth regimes \cite{WW11}. In this way, prebiotic chemistry could be more closely linked to the first steps of biological evolution.

\begin{acknowledgments}
This research is financially supported by the Belgian Federal Government
(BELSPO), the Communaut\'e fran\c caise de Belgique, and the Universit\'e Libre de Bruxelles.
\end{acknowledgments}


\begin{thebibliography}{99}

\bibitem{AG08} D. Andrieux and P. Gaspard, {\it Proc. Natl. Acad. Sci. USA} {\bf 105}, 9516 (2008).

\bibitem{J08} C. Jarzynski, {\it Proc. Natl. Acad. Sci. USA} {\bf 105}, 9451 (2008).

\bibitem{AG09} D. Andrieux and P. Gaspard, {\it J. Chem. Phys.} {\bf 130}, 014901 (2009).

\bibitem{G12} P. Gaspard, {\it Self-Organization at the Nanoscale Scale in Far-From-Equilibrium Surface Reactions and Copolymerizations}, {\tt arXiv:1203.0972} to appear in: A.~S.~Mikhailov and G.~Ertl, Editors,
Proceedings of the International Conference ``Engineering of Chemical Complexity",
Berlin Center for Studies of Complex Chemical Systems, 4-8 July 2011 (World Scientific, Singapore, 2013).

\bibitem{S44} E. Schr\"odinger, {\it What is Life?} (Cambridge University Press, Cambridge UK, 1944).

\bibitem{B79} C. H. Bennett, {\it Biosystems} {\bf 11}, 85 (1979).

\bibitem{GB06} W. J. Greenleaf and S. M. Block, {\it Science} {\bf 313}, 801 (2006).

\bibitem{Eid09} J. Eid {\it et al.}, {\it Science} {\bf 323}, 133 (2009).

\bibitem{UEAKFTP10} S. Uemura, C. Echeverr\'{\i}a Aitken, J. Korlach, B. A. Flusberg, S. W. Turner, and J. D. Puglisi, {\it Nature} {\bf 464}, 1012 (2010).

\bibitem{E92} M. Eigen, {\it Steps towards Life: A Perspective on Evolution} (Oxford University Press, Oxford, 1992).

\bibitem{ES77-78} M. Eigen and P. Schuster, {\it Naturwissenschaften} {\bf 64}, 541 (1977); {\it ibid.} {\bf 65}, 7 (1978); {\it ibid.} {\bf 65}, 341 (1978).

\bibitem{WW11} H.-J. Woo and A. Wallqvist, {\it Phys. Rev. Lett.} {\bf 106}, 060601 (2011).

\end{thebibliography}
\end{document}